\documentclass[a4paper,11pt]{article}
\usepackage{pos}
\usepackage{comment}
\usepackage{booktabs}
\usepackage{bbm}

\newcommand \mres {m_{\mathrm{res}}}
\newcommand \ml {m_l}

\newcommand \mpi {m_{\pi}}
\newcommand \barpsi {\langle\bar{\psi}\psi\rangle}

\title{Finite temperature QCD phase transition with
3 flavors of M\"obius domain wall fermions}
\ShortTitle{Three flavor QCD phase transition with MDWF}

\author*[a]{Yu Zhang}
\author[a]{Yasumichi Aoki}
\author[b,c]{Shoji Hashimoto}
\author[a]{Issaku Kanamori}
\author[b,c,d]{Takashi Kaneko}
\author[a]{Yoshifumi Nakamura}

\affiliation[a]{RIKEN Center for Computational Science,  7-1-26
	\\	Minatojima-minami-machi, Chuo-ku, Kobe, Hyogo 650-0047, Japan}

\affiliation[b]{High Energy Accelerator Research Organization (KEK), Tsukuba 305-0801, Japan}

\affiliation[c]{School of High Energy Accelerator Science, The Graduate University for Advanced Studies (Sokendai), Tsukuba 305-0801, Japan}

\affiliation[d]{Kobayashi-Maskawa Institute for the Origin of Particles and the Universe, Nagoya University, Nagoya 464–8602, Japan}

\emailAdd{yu.zhang.ey@riken.jp}

\abstract{We investigate the finite temperature QCD phase transition with three degenerate quark flavors using M\"{o}bius domain wall fermions. To explore the order of phase transition on the lower left corner of Columbia plot and if possible, to locate the critical endpoint
	we performed simulations at temperatures around 181 and 121 MeV with lattice spacing $a=0.1361(20)$~fm corresponding to temporal lattice extent $N_{\tau}=8,12$ with varying quark mass for two different volumes with aspect ratios $N_{\sigma}/N_{\tau}$ ranging from 2 to 3. By analyzing the volume and mass dependence of the chiral condensate, disconnected chiral susceptibility and Binder cumulant we find that there is a crossover at $m_q^{\mathrm{\overline {MS}}}(2\, \mathrm{GeV}) \sim 44\, \mathrm{MeV}$ for $\mathrm{T_{pc}}\sim$ 181 MeV, At temperature 121 MeV, the binder cumulant suggests a crossover at  $m_q^{\mathrm{\overline {MS}}}(2\,  \mathrm{GeV}) \sim 3.7\, \mathrm{MeV}$, 
although a study of volume dependence would be important to confirm this.}

\FullConference{%
The 39th International Symposium on Lattice Field Theory,\\
8th-13th August, 2022,\\
Rheinische Friedrich-Wilhelms-Universität Bonn, Bonn, Germany
}

\begin{document}
\maketitle

\section{Introduction}
The nature of QCD phase transition in the chiral limit is a subject of ongoing study over many decades. It depends on the number of quark flavors and their masses.  This information for $N_f=2+1$ QCD is summarized in the so-called Columbia plot. According to Pisarski and Wilczek,
the analysis of renormalization group flow in the effective $\sigma$ model with perturbative $\epsilon$ expansion  predicts that 
 the order of phase transition  in the $N_f=3$ chiral limit is expect to be first order~\cite{PhysRevD.29.338}. If this is true, the first-order phase transition gets weak away from the chiral limit and terminates at a critical endpoint of the second-order phase transition, which belongs to the 3d $Z_2$ universality class. This critical point separates the first-order and crossover transition regions.
 On the contrary, a recent study of the renormalization group flow of all couplings up to $\phi^6$ in 3d Ginzburg-Landau theory for the $N_f=3$ chiral limit claimed that the transition is second order~\cite{PhysRevD.105.L071506}. Whether the first order region exists in the $N_f=3$ light quark regime at all is still an open question, if the first-order region exists, then what is the value of the critical mass? To answer those questions, we need the nonperturbative lattice QCD simulation. 
 
 It's a hard task to pin down the location of critical mass or to determine the order of phase transition in the $N_f=3$ chiral limit using lattice QCD simulation.  Previous studies with staggered as well as Wilson and their improved fermion actions have shown that the value of the critical mass gets smaller as the lattice spacing is reduced. But the critical mass obtained from the Wilson type fermions is always larger than those obtained from the staggered type fermions~\cite{KARSCH200141,DEFORCRAND2003170, PhysRevD.54.7010, PhysRevD.91.014508,PhysRevD.96.034523,PhysRevD.95.074505}. Those indicate strong discretization scheme dependence. While a study with $\mathcal{O}(a)$-improved Wilson fermion found an upper bound of critical mass in the continuum limit $m_{\pi}^c \lesssim 110 \,\mathrm{MeV}$~\cite{PhysRevD.101.054509}, a recent study with improved staggered action (HISQ) shows that at finite lattice spacing the chiral phase transition is second order in the $N_f=3$ chiral limit~\cite{PhysRevD.105.034510}. Subsequently, a study of chiral phase transition as a function of $N_f$ with unimproved staggered fermion found a similar result that the chiral phase transition is second order in the continuum chiral limit for $N_f=3$ QCD~\cite{Cuteri:2021ikv}.  In principle, different fermion formulations should give the same result in the continuum limit, but this is not the case. The currently existing results are either from staggered or Wilson fermion which breaks chiral symmetry partially or entirely,
and it is important to use the chiral fermion formulation to further explore this problem. 

In this study, we use the 
 M\"{o}bius domain wall fermion to investivgate this $N_f=3$ chiral region. An advantage of M\"{o}bius domain wall fermion is that it possesses accurate chiral symmetry even at finite lattice spacing when the fifth dimension $L_s$ is sufficiently large.
   \section{Simulation Setup}
We perform $N_f=3$ QCD simulations with the tree-level improved Symanzik gauge action and M\"{o}bius domain wall fermion action~\cite{Nakamura:2022abk} by using the optimized code set Grid which adapts to the Fugaku CPU A64FX~\cite{Meyer:2019gbz}.  We choose the fixed gauge coupling $\beta=4.0$ which corresponds to the lattice spacing $a=0.1361(20)$ fm, which is determined from the Wilson flow $t_0$ and matches with $N_f=2+1$ QCD physical point result~\cite{Borsanyi:2012zs}. The calculation is carried out on lattices of size $N_{\sigma}^3 \times 8$ with $N_{\sigma} = 16, 24$, and $24^3\times 12$ corresponding to temperatures of 181 and 121 MeV, respectively. We simulate 21 quark masses which are in the interval $am \in \left[ 0, 0.2\right]$ for $N_{\tau}=8$ lattices, and 26 quark masses which are in the interval $am \in \left[ -0.006, 0.1\right]$ for $24^3\times12$ lattices. In addition to these finite temperature ensembles, we also generated ensembles at $\beta=4.0$ with lattices of size $24^3\times 48$ and $12^3\times24$ for several different quark masses. We use these zero temperature ensembles to determine the residual mass, which characterize the residual chiral symmetry breaking due to finite $L_s$, and chiral condensate, which can be used to remove the UV divergence of finite temperature chiral condensate. $L_s$ is 16 for all lattices. 
 \section{Observables}
\subsection{Residual mass}
The finite fifth dimension of domain wall fermion allows the mixing between the left and right-handed zero modes, which results in the residual chiral symmetry breaking. To leading order in an expansion in lattice spacing, this residual chiral symmetry breaking can be characterized by the residual mass $\mres$, which is measured through
\begin{equation}
  \label{eq:ratio}
 R(t) = \frac{\langle \sum_{\vec x}J_{5q}^{a}(\Vec{x},t)\,\pi^{a}(\vec{0},0) \rangle}{\langle \sum_{\vec x}J_5^a(\vec{x},t)\,\pi^a(\vec{0},0)\rangle} \,,
\end{equation}
where $J_5^a$ is the pseudoscalar density which is constructed by the quark fields on the boundary of the fifth dimension and $J_{5q}^a$ is the pseudoscalar density built from the quark fields at $L_s/2$ and $L_s/2-1$. $\mres$ at a given input quark mass is obtained by fitting $R(t)$ to a constant at large source-sink separation $t$, where it shows a good plateau and only pions contribute to the correlators.

\begin{figure}[!htp]
	\centering
	\includegraphics[width=0.35\textwidth]{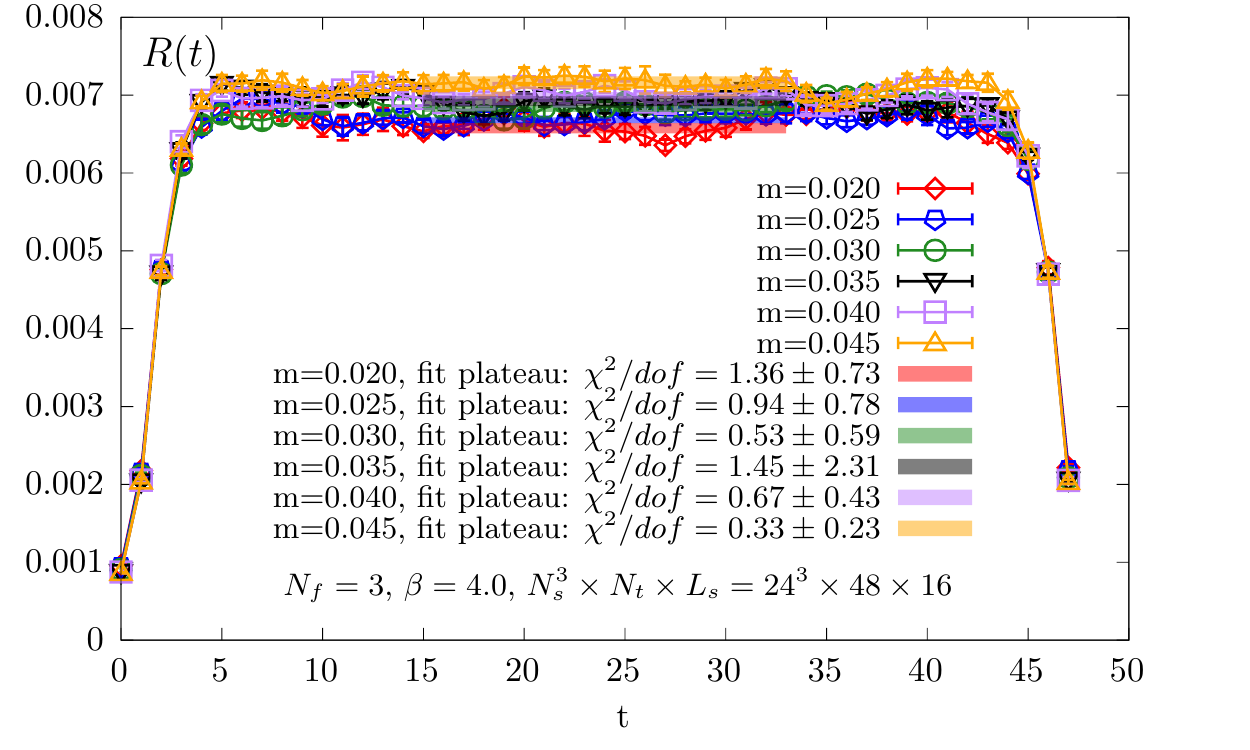}~
	\includegraphics[width=0.35\textwidth]{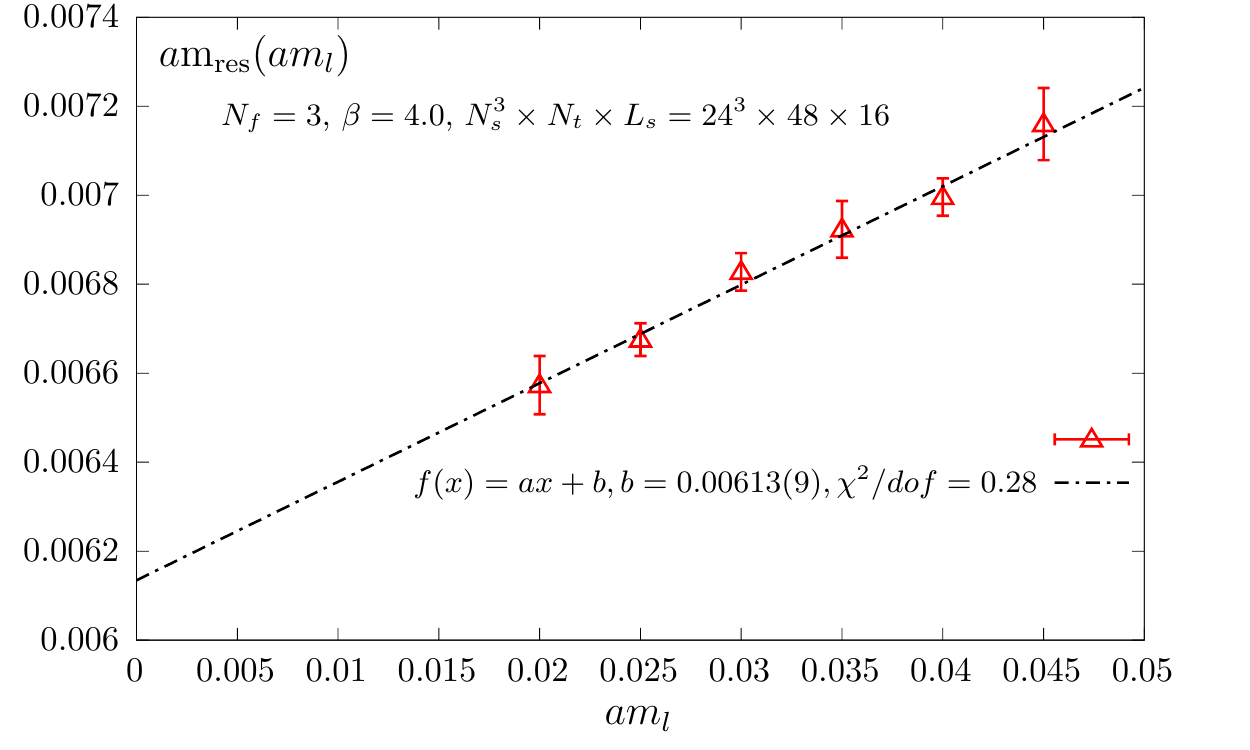}~
		\includegraphics[width=0.35\textwidth]{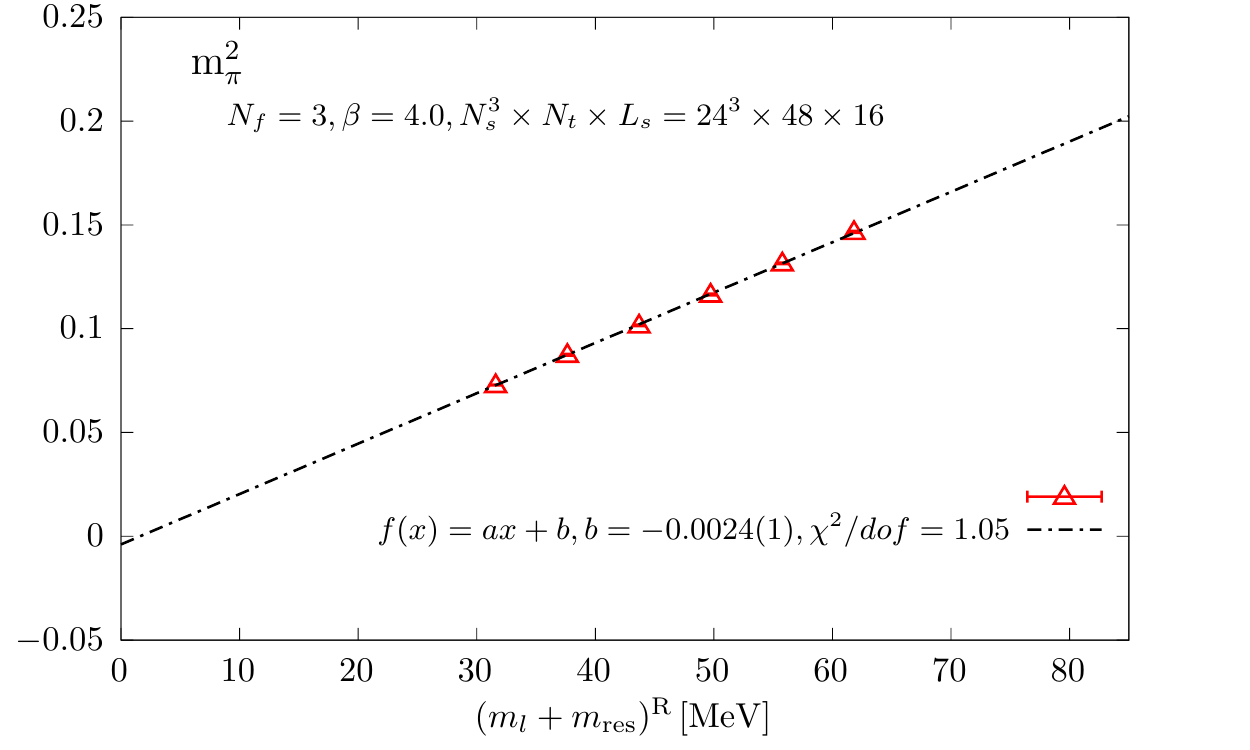}
	\caption{Left: The ratio of boundary-midpoint to the boundary-boundary pseudoscalar correlators as a function of source-sink separation $t$ for each quark mass on $24^3\times 48$ lattices at $\beta=4.0$ with $L_s=16$. The horizontal bands represent the fit to a constant over the range $15\le t \le 33$, which gives residual mass. Middle: The residual mass as a function of the input quark mass. The dashed line is a linear fit to the input quark mass. Right: The pion mass squared as calculated from the pion correlator $\langle \pi^a(x)\pi^a(0)\rangle$versus the renormalized quark mass in physical units. The dashed line is a linear fit to the renormalized quark mass which has been converted into the $\overline{\mathrm {MS}}$ scheme and expressed in the unit of $\mathrm{MeV}$.} 
	\label{fig:R_mres}
\end{figure}

The left plot of~\autoref{fig:R_mres} shows the results of $R(t)$ as a function of $t$ for  $24^3\times 48\times 16$ lattices at $\beta=4.0$ with different quark masses. The horizontal bands represent the fit to a constant over the range $15\le t \le 33$ with jackknife errors for every quark mass, to determine $a\mres(a\ml)$.
The middle plot of~\autoref{fig:R_mres} shows the result of $a\mres(a\ml)$ as a function of input quark mass $a\ml$. We performed a linear fit (dashed lines), which gives a good description of the data. This linear dependence on the input quark mass $a\ml$ is understood in the literature to be a lattice artifact where $a\mres(a\ml) = a\mres(a\ml=0) (1+\mathcal{O}(am_la^2\Lambda)$~\cite{Sharpe:2007yd}, the common way for dealing with this lattice artifact and define a mass independent $a\mres$ is an extrapolation of $a\mres(a\ml)$ to the zero input quark mass limit. It is determined as 
\begin{align}\label{eq:mres_val}
a\mres \equiv a\mres(a\ml=0) = 0.00613(9)\,.
\end{align}
The residual mass acts as an additive shift to the input quark mass, giving the total quark mass $m = \ml + \mres$. Thus, the chiral limit is defined as $\ml + \mres =0$. The above definition of $\mres$ makes $\mpi$ vanishes in the chiral limit. The right plot of~\autoref{fig:R_mres} shows the result of $\mpi^2$ as a function of renormalized total quark mass in the $\overline{\mathrm {MS}}$ scheme at a scale of 2 GeV and physical scale which is given by 
$m^{\overline{\mathrm {MS}}}(2\, \mathrm{GeV}) \equiv  (\ml+\mres)^{R} = Z_m^{\overline{\mathrm {MS}}}(2\, \mathrm{GeV}) \cdot a^{-1} \cdot (a\ml+a\mres)$.
The value of $\mres$ is determined in \autoref{eq:mres_val}. $Z_m^{\overline{\mathrm {MS}}}(2\, \mathrm{GeV})$ is the mass renormalization constant which is determined by applying the NNNLO running. It takes the value of 0.834348 for $\beta=4.0$~\cite{Aoki:2021kbh}.
The dashed line is a linear fit as a function of the renormalized quark mass. $\mpi^2$ is close to zero in the chiral limit but not exactly extrapolated to zero. This is probably due to ignored chiral logarithm term and $\mathcal{O}(a^2)$ effect. The demonstration of $\mpi^2$ vanishing in the chiral limit is a critical component for establishing the good chiral properties of the domain wall fermion.
\subsection{Chiral condensate}
The chiral condensate $\langle \bar{\psi}{\psi} \rangle$ is the order parameter of the QCD chiral phase transition. It vanishes in the chirally symmetric phase and remains nonzero in the chirally broken phase. It is defined as
\begin{align}\label{eq:pbp}
\barpsi = \frac{T}{V}\frac{\partial \ln Z}{\partial m} = \frac{1}{N_{\sigma}^3N_{\tau}}\left\langle \mathrm{Tr} M^{-1}\right\rangle\,,
\end{align}
where $Z$ is the partition function, $T$ is temperature and $V$ is spatial volume, and $M$ is the Dirac matrix.
At finite quark mass, $\langle \bar{\psi}{\psi} \rangle$ requires both additive and multiplicative renormalizations. The former usually arises from the $m_l/a^2$ divergence to the chiral condensate. For domain wall fermion with finite $L_s$, it's more complicated, as the chiral condensate receives contributions from the energy scale up to the cutoff scale, which is much larger than those for $\mres$ representing the effect of $\chi_{SB}$, as discussed in~\cite{Sharpe:2007yd}. The chiral condensate behaves as
\begin{align}\label{eq:pbp}
\barpsi|_{\mathrm{DWF}} \sim \frac{\ml + x\mres}{a^2} + \barpsi|_{\mathrm{cont}} + ...\,,
\end{align}
where $x$ is unknown. Instead of $x=1$, one expects $x=\mathcal{O}(1)$. If we extrapolate to the chiral limit $\ml+\mres=0$, there still remain some UV divergent pieces, which behave as $\frac{(x-1)\mres}{a^2}$. This unwanted contribution to $\barpsi$ can only be controlled by increasing $L_s\to \infty$. On the other hand, since these UV divergent terms $\frac{\ml + x\mres}{a^2}$ have no temperature dependence, they can be removed by subtracting zero temperature $\barpsi$ at the same input quark mass $\ml$:
	$\barpsi^{\mathrm{T}>0} - \barpsi^{\mathrm{T}=0}$.
The remaining multiplicative divergence of chiral condensate can be removed by multiplying
$Z_s^{\overline{\mathrm {MS}}}(2\, \mathrm{GeV})$, which is equal to $1/Z_m^{\overline{\mathrm {MS}}}(2\, \mathrm{GeV})$. This converts the result into the $\overline{\mathrm {MS}}$ renormalization scheme at a scale $\mu=2\,\mathrm{GeV}$:
\begin{align}\label{eq:renorm_pbp}
[\barpsi^{\mathrm{T}>0} - \barpsi^{\mathrm{T}=0}]^{\overline{\mathrm {MS}}}(2\,\mathrm{GeV}) = \frac{\barpsi^{\mathrm{T}>0} - \barpsi^{\mathrm{T}=0}}{Z_m^{\overline{\mathrm {MS}}}(2\,\mathrm{GeV})}\,.
\end{align}

\begin{figure}[!htp]
	\centering
	\includegraphics[width=0.45\textwidth, height=0.215\textheight]{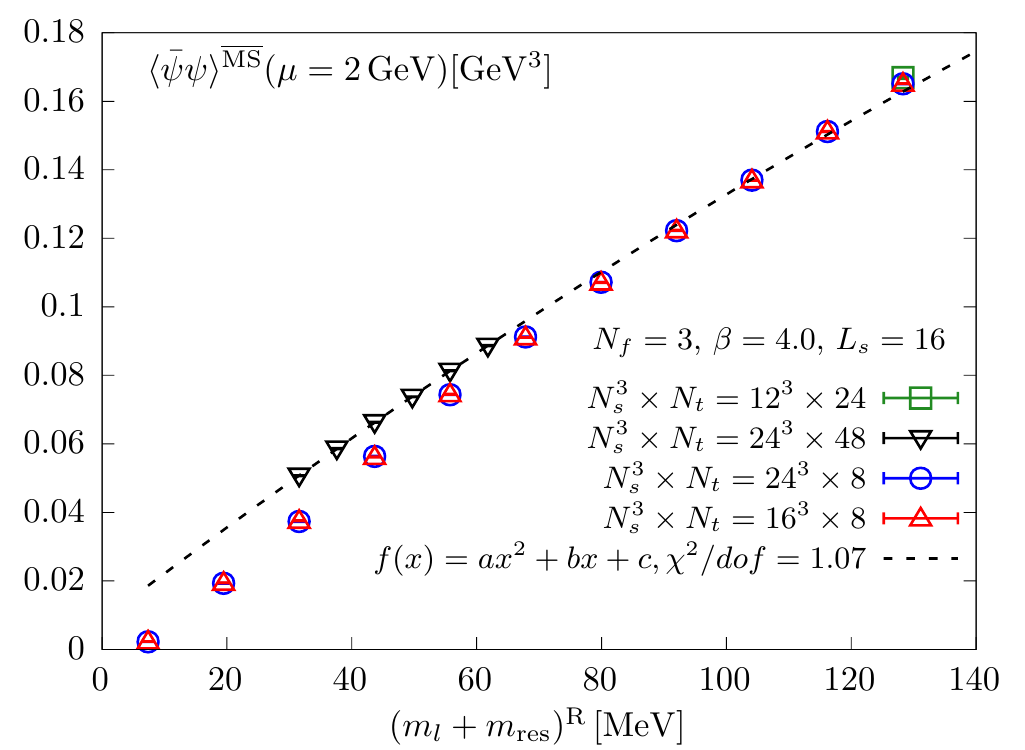}
	\includegraphics[width=0.45\textwidth, height=0.215\textheight]{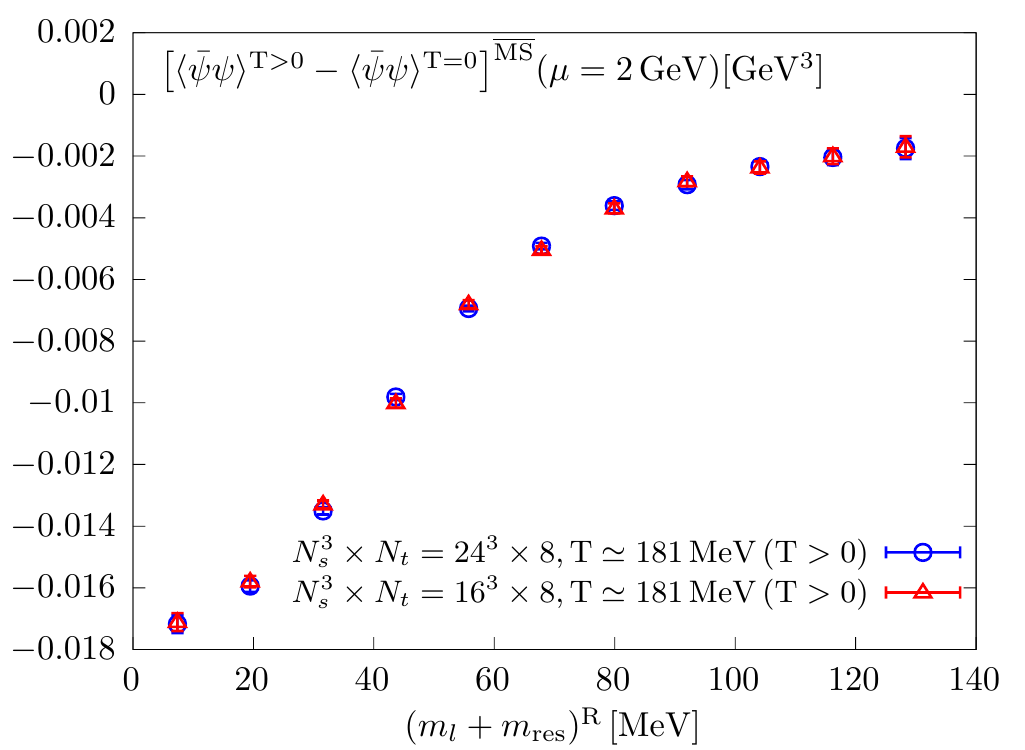}
	\caption{Left: Renormalized chiral condensate on finite temperature lattices  $16^3\times8$, $24^3\times8$ and zero temperature lattices $12^3\times24$, $24^3\times48$ for $\beta=4.0$ as a function of renormalized quark mass. The dashed line is the linear extrapolation to the renormalized quark mass for zero temperature chiral condensate. Right:  The subtracted chiral condensate as a function of renormalized quark mass for $16^3\times8$ and $24^3\times8$ lattices.
	All results are converted into the $\overline{\mathrm {MS}}$ renormalization scheme. The vertical axis is in the unit of $\mathrm{GeV}^3$ and the horizontal axis is in the unit of $\mathrm{MeV}$.} 
	\label{fig:Nt8_pbp}
\end{figure}

For the finite temperature ensembles, we measure chiral condensate using 10 stochastic noise vectors on every tenth trajectory. The use of multiple stochastic noise vectors allows us to estimate the fluctuations of $\bar{\psi}\psi$ such as disconnected chiral susceptibility in an unbiased way.
The left plot of~\autoref{fig:Nt8_pbp} shows $\langle\bar{\psi}\psi\rangle^{\overline{\mathrm {MS}}}(2\,\mathrm{GeV})$ versus renormalized quark mass both are converted into $\overline{\mathrm {MS}}$ scheme and expressed in the physical unit for zero temperature ensembles and finite temperature ensembles at $N_{\tau}=8$ for two different volumes with aspect ratios $N_{\sigma}/N_{\tau}$ ranging from 2 to 3 at $\beta=4.0$. The result of $\langle\bar{\psi}\psi\rangle^{\overline{\mathrm {MS}}}(2\,\mathrm{GeV})$ for $N_{\tau}=8$ goes to zero before reaching the chiral limit. This is due to the remnant UV divergence $\frac{(x-1)\mres}{a^2}$. To get rid of it, we perform a quadratic extrapolation of zero temperature $\langle\bar{\psi}\psi\rangle^{\overline{\mathrm {MS}}}(2\,\mathrm{GeV})$ to the renormalized quark mass, which is represented by the dashed line and then subtract from the finite temperature $\langle\bar{\psi}\psi\rangle^{\overline{\mathrm {MS}}}(2\,\mathrm{GeV})$, i.e. $[\barpsi^{\mathrm{T}>0} - \barpsi^{\mathrm{T}=0}]^{\overline{\mathrm {MS}}}(2\,\mathrm{GeV})$,  as shown in 
the right plot of~\autoref{fig:Nt8_pbp}.
Here all the divergences of chiral condensate have been removed. We observe no volume dependence between lattices of size $24^3\times8$  and $16^3\times8$. This implies a smooth crossover transition for $\mathrm{T}\sim 181\,\mathrm{MeV}$. 

\begin{figure}[!htp]
	\centering
	\includegraphics[width=0.45\textwidth, height=0.215\textheight]{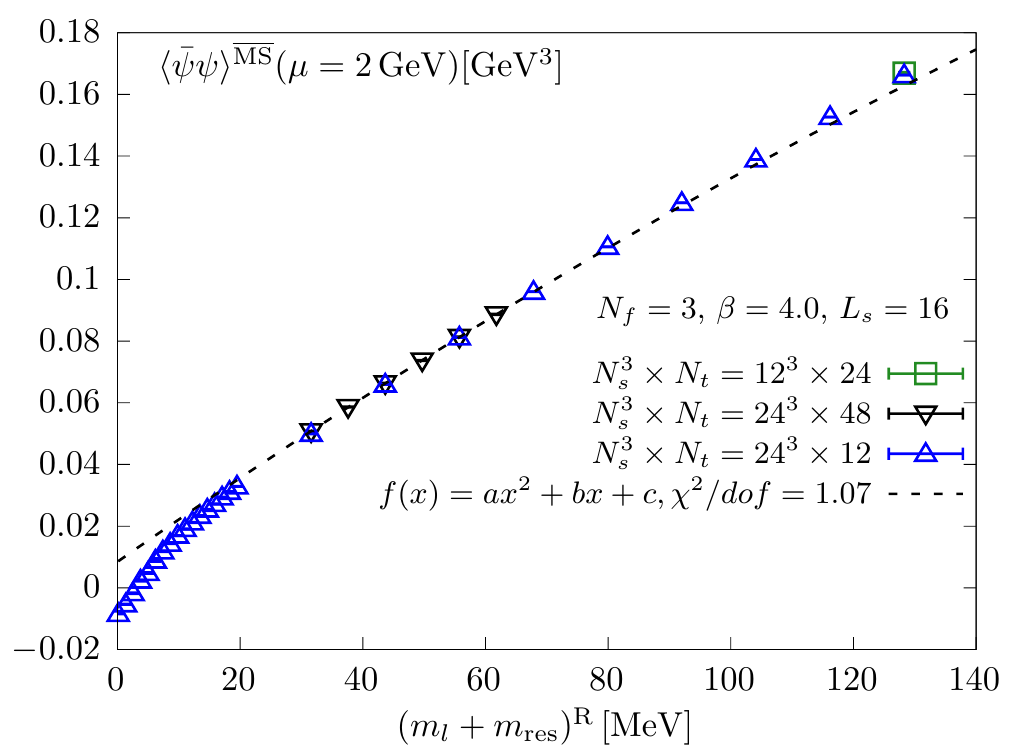}
	\includegraphics[width=0.45\textwidth, height=0.215\textheight]{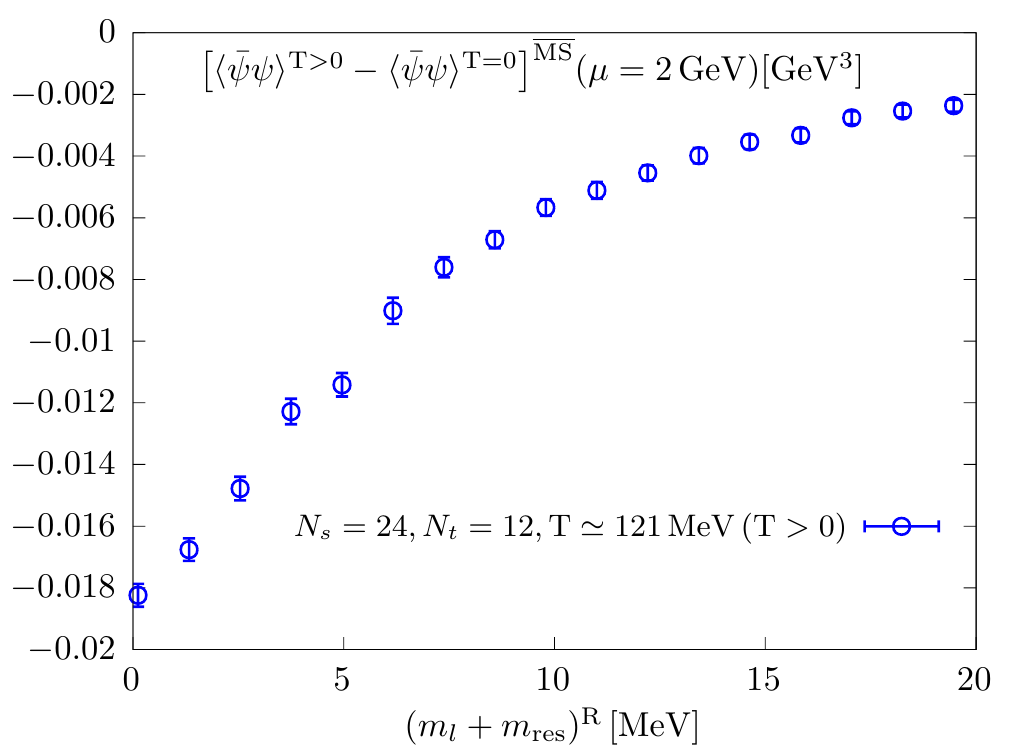}
	\caption{Same as~\autoref{fig:Nt8_pbp} but the finite temperature chiral condensate is measured on $24^3\times 12$ lattices. } 
	\label{fig:Nt12_pbp}
\end{figure}

We show similar plots but for $N_{\tau}=12$ finite temperature ensembles in~\autoref{fig:Nt12_pbp}. In the left plot of~\autoref{fig:Nt12_pbp}, the negative chiral condensate result in the chiral limit is due to the remnant additive divergence. The subtracted chiral condensate as a function of  renormalized quark mass is shown in the right plot of~\autoref{fig:Nt12_pbp}. Since we only have one volume for $N_{\tau}=12$, it's hard to determine whether this is a crossover or true phase transition from the behavior of chiral condensate. The inflection point is roughly around 3$\sim$7 MeV for $\mathrm{T}\sim 121\,\mathrm{MeV}$. 
\subsection{Disconnected chiral susceptibility}
In order to locate the inflection point of chiral condensate precisely, we measure the disconnected part of chiral susceptibility $\chi_{\mathrm{disc}}$. It does not suffer from additive divergence but requires multiplicative renormalization~\cite{PhysRevD.85.054503}. The multiplicative divergence can be removed by renormalize $\chi_{\mathrm{disc}}$ to the $\overline{\mathrm {MS}}$ scheme at a scale of 2 GeV with $\left(Z_m^{\overline{\mathrm {MS}}}(2\,\mathrm{GeV})\right)^{-2}$ as 
\begin{align}\label{eq:renorm_chi_disc}
\chi_{\mathrm{disc}}^{\overline{\mathrm {MS}}}(2\,\mathrm{GeV}) = \left(\frac{1}{Z_m^{\overline{\mathrm {MS}}}(2\,\mathrm{GeV})}\right)^2\left(\frac{1}{N_{\sigma}^3N_{\tau}}\left(\left\langle (\mathrm{Tr} M^{-1})^2\right\rangle  -\left\langle \mathrm{Tr} M^{-1}\right\rangle^2\right) \right)\,.
\end{align}
$\chi_{\mathrm{disc}}$ describes the fluctuation of $\barpsi$ and shows a peak at the inflection point of $\barpsi$.

\begin{figure}[!htp]
	\centering
	\includegraphics[width=0.45\textwidth, height=0.215\textheight]{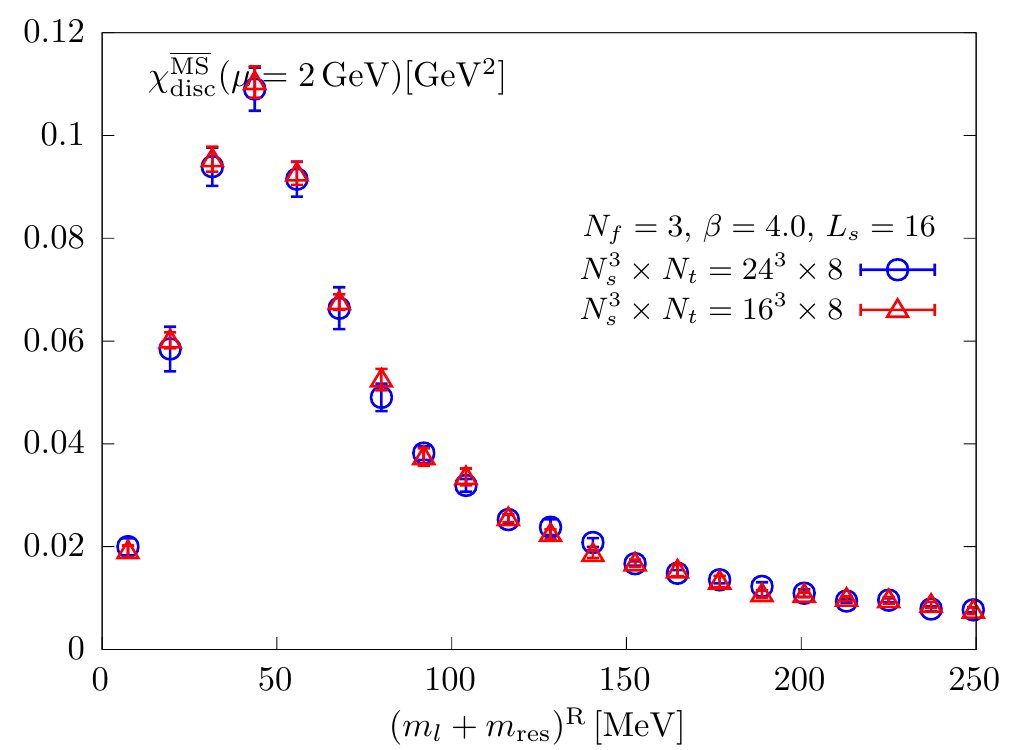}
	\includegraphics[width=0.45\textwidth, height=0.222\textheight]{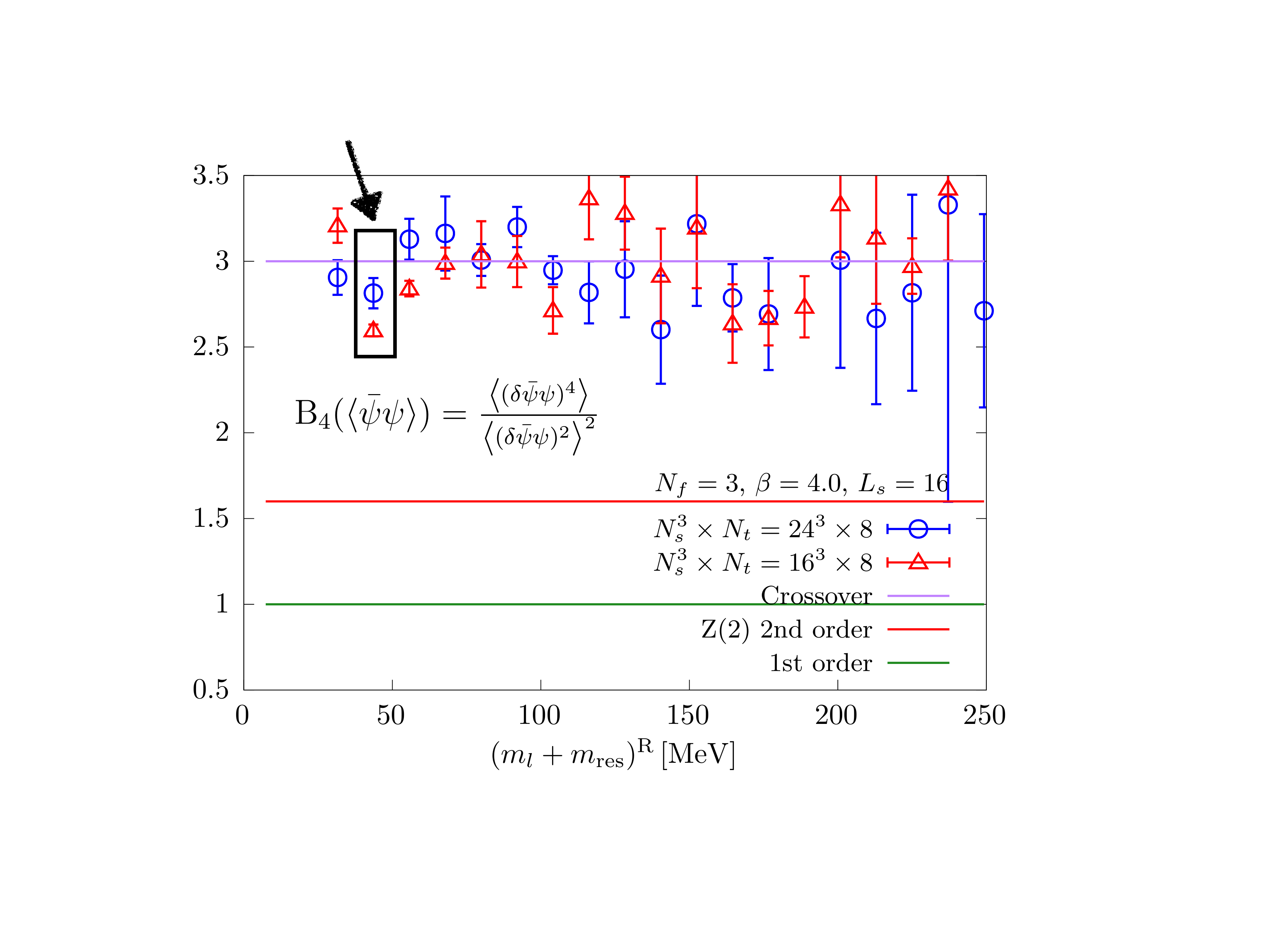}
	\caption{Disconnected chiral susceptibility $\chi_{\mathrm{disc}}^{\overline{\mathrm {MS}}}(2\,\mathrm{GeV})$ renormalized in $\overline{\mathrm {MS}}$ scheme (left) and Binder cumulant of chiral condensate (right) as a function of renormalized quark mass for $24^3\times 8$ and $16^3\times 8$ lattices with $L_s=16$ at $\beta=4.0$. The black rectangle marks the result of $B_4(\bar\psi \psi)$ at transition point of this fixed temperature.  } 
	\label{fig:Nt8_chi_disc}
\end{figure}

\begin{figure}[!htp]
	\centering
	\includegraphics[width=0.45\textwidth, height=0.215\textheight]{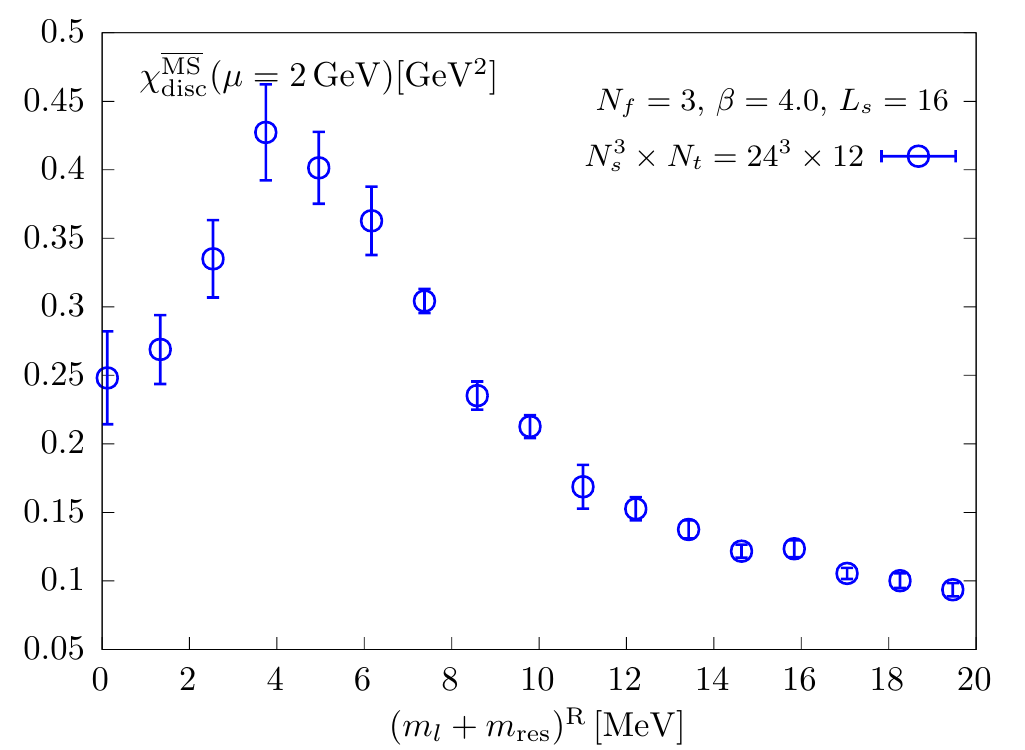}
	\includegraphics[width=0.45\textwidth, height=0.222\textheight]{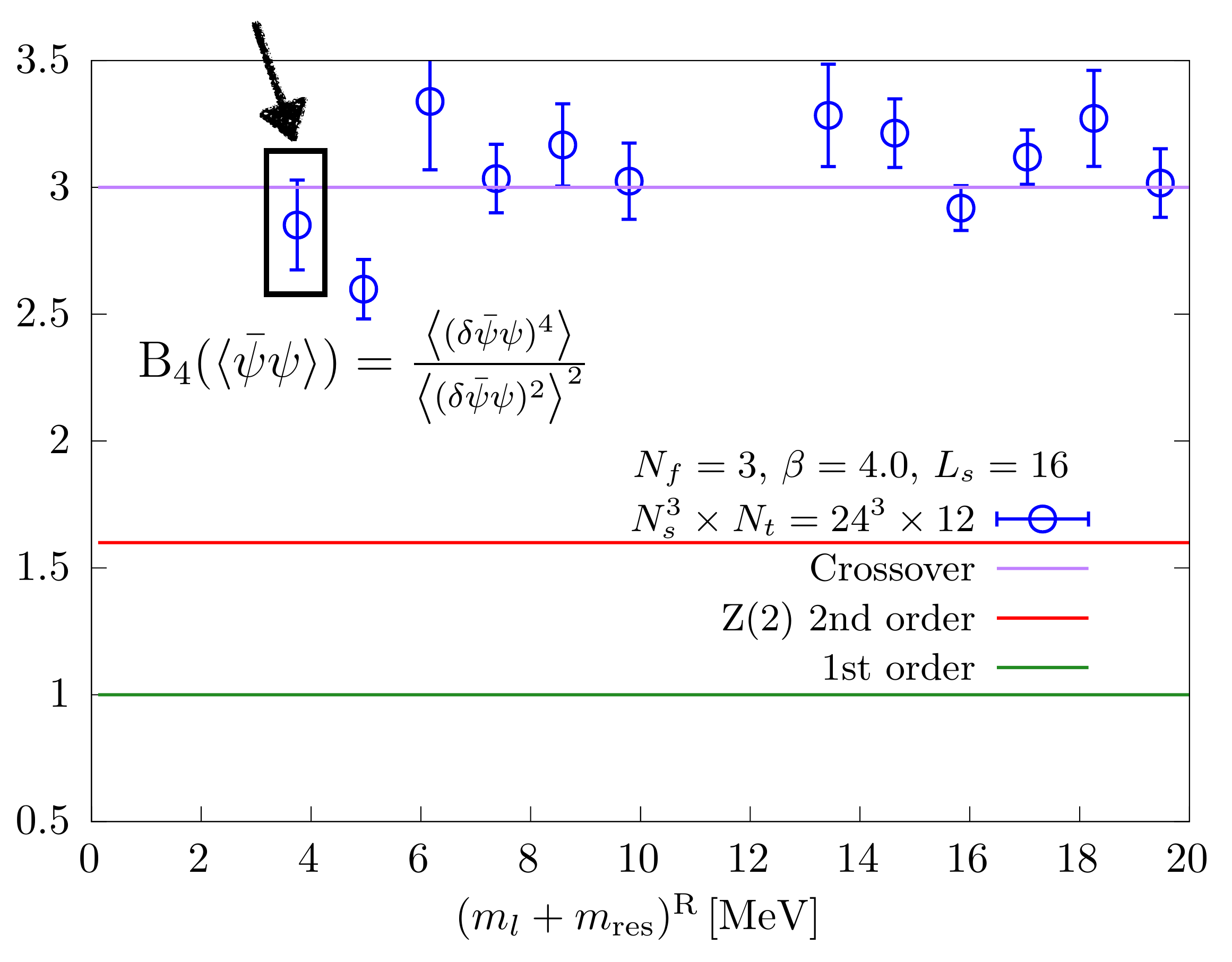}
	%
	\caption{Same as~\autoref{fig:Nt8_chi_disc} but $\chi_{\mathrm{disc}}^{\overline{\mathrm {MS}}}(2\,\mathrm{GeV})$ and $B_4(\bar\psi \psi)$ are measured on $24^3\times 12$ lattices. }
	\label{fig:Nt12_chi_disc}
\end{figure}

The left plot of~\autoref{fig:Nt8_chi_disc} shows the disconnected chiral susceptibility as a function of renormalized quark mass for lattices of size $N_{\sigma}^3 \times 8$ with $N_{\sigma} = 16, 24$. This susceptibility shows a pronounced peak at around 44 MeV. For a real phase transition, the peak height of $\chi_{\mathrm{disc}}$ will increase as increase volume, while the data shows no volume dependence. It implies an analytic crossover and the pseudo critical mass is around 44 MeV for T$\sim$181 MeV.  
In the left plot of~\autoref{fig:Nt12_chi_disc}, we show $\chi_{\mathrm{disc}}$ as a function of renormalized quark mass for $24^3\times12$ lattices, one can see that the inflection point is around 3.7 MeV where $\chi_{\mathrm{disc}}$ shows a peak. Since we only have one volume, we cannot tell the order of phase transition from the behavior of $\chi_{\mathrm{disc}}$.
\subsection{Binder cumulant}
To determine the order of the phase transition at the inflection point of $\barpsi$, we measured the Binder cumulant of chiral condensate defined as
\begin{align}\label{eq:B4}
B_4(\bar\psi \psi) = \frac{\left\langle (\delta \bar\psi \psi)^4\right\rangle}{\left\langle (\delta\bar\psi\psi)^2\right\rangle^2},\quad \delta\bar\psi \psi = \bar\psi \psi - \barpsi\,.
\end{align}
To estimate $B_4(\bar\psi \psi)$ in an unbiased way, we choose multiple stochastic noise vectors and use the chiral condensate which comes from the different stochastic noise vectors to calculate the powers of chiral condensate~\cite{Cuteri:2021ikv}. The value of $B_4(\bar\psi \psi)$ can be used to distinguish the order of phase transition. In the thermodynamic limit, $B_4(\bar\psi \psi)=1$ corresponds to a first-order phase transition,  $B_4(\bar\psi \psi)=3$ to an analytic crossover, and $B_4(\bar\psi \psi)= 1.604$ to the second order phase transition with 3-dimensional Z(2) universality class. However, at finite volume, the result of $B_4(\bar\psi \psi)$ has volume dependence for the first-order phase transition and crossover. It will approach the corresponding universal value in the infinite volume limit, while the value of $B_4(\bar\psi \psi)$ does not change at a second-order transition point for different lattices i.e., it is scale invariant. Our goal is to determine the order of phase transition for the transition point at a fixed temperature.

We show the result of $B_4(\bar\psi \psi)$ calculated on $24^3\times 8$ and $16^3\times 8$ lattices at $\beta=4.0$ for different quark masses in the right panel of~\autoref{fig:Nt8_chi_disc}. Here we only focus on the transition point at this fixed temperature which is determined from the peak location of $\chi_{\mathrm {disc}}$ shown in the left plot of~\autoref{fig:Nt8_chi_disc} and marked by a black rectangle. It seems that $B_4(\bar\psi \psi)$ approaches 3 as you increase the volume to the thermodynamic limit. This is expected for a crossover transition at $m_q^{\mathrm{\overline {MS}}}(2\, \mathrm{GeV}) \sim 44\, \mathrm{MeV}$. This is consistent with what we have found from the volume independence of $\barpsi$ and $\chi_{\mathrm {disc}}$ as shown in the right plot of~\autoref{fig:Nt8_pbp} and left plot of~\autoref{fig:Nt8_chi_disc}.

In the right plot of~\autoref{fig:Nt12_chi_disc}, we show a similar plot as the right plot of~\autoref{fig:Nt8_chi_disc} but for $24^3\times 12$ lattices. The black rectangle marks the result of $B_4(\bar\psi \psi)$ at the pseudo critical or critical quark mass for temperature 121 MeV. It's close to 3. It seems to be a crossover, but another larger lattice for $N_{\tau}=12$ would be important to confirm this.
\section{Summary and Outlook}
We have shown the first study of $N_f=3$ QCD phase transition in the chiral regime using chiral fermion formulation, M\"{o}bius domain wall fermions.  We have evaluated the chiral condensate, disconnected chiral susceptibility, and Binder cumulant for fixed temperature 121 MeV with lattices $24^3\times 12$ and 181 MeV with lattices $16^3\times 8$ and $24^3\times8$, respectively.  
For each lattice,  we have a variety of quark masses.  
We have also examined zero temperature chiral condensate which is used to remove the additive divergence for the finite temperature chiral condensate.

We observed that at temperature 181 MeV, the chiral condensate and disconnected susceptibility has no volume dependence for all the quark masses, and $\chi_{\mathrm{disc}}$ exhibits a peak at $m_q^{\mathrm{\overline {MS}}}(2\, \mathrm{GeV}) \sim 44\, \mathrm{MeV}$, and further the result of $B_4(\bar\psi \psi)$ close to 3 towards the thermodynamic limit at transition point $m_q^{\mathrm{\overline {MS}}}(2\, \mathrm{GeV}) \sim 44\, \mathrm{MeV}$. These all indicate that it's a crossover phase transition at $m_q^{\mathrm{\overline {MS}}}(2\, \mathrm{GeV}) \sim 44\, \mathrm{MeV}$ for temperature 181 MeV. 
For temperature 121 MeV, $\chi_{\mathrm{disc}}$ shows a peak at $m_q^{\mathrm{\overline {MS}}}(2\, \mathrm{GeV}) \sim 3.7\, \mathrm{MeV}$ and the result of $B_4(\bar\psi \psi)$ is close to 3 at this transition point. This means it might be a crossover, but we need another larger lattices to confirm this, currently the simulation of $36^3\times12$ lattices is underway. If this is indeed the crossover, then the critical mass should be smaller than 3.7 MeV and this is consistent with the findings from the Wilson and staggered fermion. 

We are trying to study the residual chiral symmetry breaking effect by increasing $L_s$. Currently, the simulation for $L_s=32$ is underway. And we want to investigate the lower temperature, which is exploring the lighter quark mass. For that, we are starting the simulation for $N_t=14$.
\section{Acknowledgments}
This work used the computational resources of Supercomputer Fugaku provided by the RIKEN Center for Computational Science through HPCI project hp210032 and Usability Research ra000001 as well as Wisteria/BDEC-01 Supercomputer System at Tokyo University/JCAHPC through HPCI project hp220108 and Ito supercomputer at Kyushu University through HPCI project hp190124 and hp200050 and also the Hokusai BigWaterfall at RIKEN. This work is also supported in part by JSPS KAKENHI grant No 20H01907 and 21H01085.  We acknowledge the Grid Lattice QCD framework\footnote[1]{https://github.com/paboyle/Grid} and its extension for A64FX processors~\cite{Meyer:2019gbz} which is used to generate the QCD configurations for this study. We thank N. Meyer and T. Wettig for discussions on the use of Grid for A64FX. For the measurement, we used Bridge++~\cite{Ueda:2014rya} and Hadrons code~\cite{antonin_portelli_2020_4293902}. 

\bibliographystyle{JHEP.bst}
\bibliography{ref.bib}

\end{document}